\documentclass[conference]{IEEEtran}
\usepackage{cite}
\usepackage{amsmath,amssymb,amsfonts}
\usepackage{algorithmic}
\usepackage{graphicx}
\usepackage{textcomp}
\usepackage{xcolor}
\usepackage{fancyhdr}
\usepackage[hyphens]{url}

\usepackage{subfigure}
\usepackage{tikz}
\newcommand {\circled}[1]{\lower.7ex\hbox{\tikz\draw (0pt, 0pt)%
		circle (.5em) node {\makebox[1em][c]{\small #1}};}}

\newcommand{\tabincell}[2]{\begin{tabular}{@{}#1@{}}#2\end{tabular}}
\def\BibTeX{{\rm B\kern-.05em{\sc i\kern-.025em b}\kern-.08em
		T\kern-.1667em\lower.7ex\hbox{E}\kern-.125emX}}

\title{A Write-Friendly and Fast-Recovery Scheme for Security Metadata in NVM} 
\author{
	{\rm Jianming Huang, Yu Hua}
	\\
	Huazhong University of Science and Technology \\
}
\begin{document}
	\maketitle
	\pagestyle{plain}

%%%%%% -- PAPER CONTENT STARTS-- %%%%%%%%
	
%-------------------------------------------------------------------------------
\begin{abstract}
%-------------------------------------------------------------------------------
Non-Volatile Memories (NVMs) have attracted the attentions of academia and industry, which is expected to become the next-generation memory. However, due to the nonvolatile property, NVMs become vulnerable to attacks and require security mechanisms, e.g., counter mode encryption and integrity tree, which introduce the security metadata. NVMs promise to recover these security metadata after a system crash, including the counter and integrity tree. However, unlike merkle tree reconstructed from user data, recovering SGX integrity tree (SIT) has to address the challenges from unique top-down hierarchical dependency. Moreover, writing overhead and recovery time are important metrics for evaluating persistent memory system due to the high costs of NVM writes and IT downtime. How to recover the security metadata, i.e., counter blocks and integrity tree nodes, with low write overhead and short recovery time, becomes much important.
		
To provide a fast recovery scheme with low write overhead, we propose STAR, a cost-efficient scheme for recovering counter blocks and SGX integrity tree nodes after crashes. For fast recovery and verification, STAR synergizes the MAC and correct data, uses bitmap lines in ADR to indicate the location of stale node and constructs a cached merkle tree to verify the correctness of the recovery process. Moreover, STAR uses a multi-layer index to speed up the recovery process. STAR also allows different configurations to meet adaptive requirements for write overhead and recovery time.  Our evaluation results show that the proposed STAR reduces the number of memory writes by up to 87\% compared with state-of-the-art work, Anubis, which needs extra 1x memory writes. For a 4MB security metadata cache, STAR needs 0.039s/0.023s/0.004s in three different configurations to recover the metadata cache while Anubis needs 0.020s.
		
\end{abstract}

%\input{introduction.tex}
%\vspace{-0.1cm}
%\input{background.tex}
%\vspace{-0.1cm}
%\input{motivation.tex}
%\input{design.tex}
%\vspace{-0.1cm}
%\input{methodology.tex}
%\vspace{-0.1cm}
%\input{experiment.tex}
%\vspace{-0.1cm}
%\input{discussion.tex}
%\input{related.tex}
%\vspace{-0.3cm}
%\input{conclusion.tex}
%\vspace{-0.3cm}
\section{Introduction}
\label{section 1}
Non-Volatile Memory (NVM) is promising to become the main devices of next-generation memory systems, due to high density, near-zero standby power, non-volatile and byte-addressable features. NVM however suffers from some drawbacks, such as limited cell endurance~\cite{yang2013memristive,LeeIMB09,QureshiKFSLA09,ZhouZYZ09} and asymmetric read and write latencies~\cite{YueZ13,SchechterLSB10}. More importantly, NVM suffers from more severe security vulnerabilities than DRAM due to its non-volatile feature. After physically stealing DIMM, an attacker can easily read the contents from another computer due to retaining data after power off in NVM. To protect the user data from attacks, an intuitive solution is to leverage data encryption and integrity verification, which introduces extra security metadata, i.e., counter blocks and integrity tree nodes. In general, recovering these metadata from system crashes is important to NVM~\cite{ZubairA19} to efficiently support further execution of original applications. How to recover these security metadata in the persistent memory system has been widely discussed in recent works~\cite{ZubairA19,YeHA18,YangLCMS19,AwadYSNZ19}.

Due to hiding the decryption latency, counter mode encryption scheme (CME)~\cite{lipmaa2000ctr} for secure memory systems~\cite{YoungNQ15,AwadMHSH16,SwamiRM16,Zuo0ZZG18} becomes more efficient via one-time padding than direct encryption via AES. However, in practice, data encryption becomes insufficient to ensure the security of persistent memory, since attackers can modify the encrypted data/counters without authorization. The persistent system obtains useless plaintext by decrypting these modified ciphertext data/counters. To detect the unauthorized modifications, Message Authentication Code (MAC) is used to associate data with their counters. A MAC is stored with a data line. Any modification in a counter, data or MAC can be detected by comparing the stored MAC with computed one. Unfortunately, attackers can replace the new data, counter and MAC with old tuple, which is called data replace attack. To detect this attack, an integrity tree is introduced into memory systems~\cite{RogersCPS07,ZubairA19,AwadYSNZ19}. The counter blocks are hashed to generate the MACs that are further hashed iteratively until generating a root node that is stored on-chip non-volatile register, which is called Bonsai Merkle Tree (BMT)~\cite{RogersCPS07}. Unlike BMT, there is another integrity tree, i.e., SGX integrity tree (SIT)~\cite{CostanD16,TaassoriSB18}. Each node in a traditional SIT consists of 8 counters and one MAC, and the MAC is generated by hashing the 8 counters and one corresponding counter in the father node (detailed in Section~\ref{section 2.3}). SIT updates MACs in each node in parallel, while BMT updates MACs in the same branch sequentially.

The security metadata generally include both counter blocks and integrity tree nodes, which need to be recoverable to ensure the system security after system recovery. Persistent memory recovery requires short recovery time and low run-time overhead, especially in terms of the number of memory writes. In Amazon’s cloud system, the downtime costs are up to 2 million dollars per minute~\cite{Amazon} and the average costs of IT downtime are 5,600 dollars per minute~\cite{ITDownTime}. Long recovery time makes the recovery scheme inefficient and the persistent data fail to be recovered eventually~\cite{AwadYSNZ19}. The naive recovery approach is to persist every changed node in metadata cache, i.e., strict persistence scheme~\cite{ZubairA19,AwadYSNZ19}. The applications can run immediately after system reboot without recovery process in the strict persistence scheme. However, this scheme causes all the integrity tree nodes in the branch (even tens of levels) from leaf node to the tree root changed with user data to be written. In this case, all changed nodes have to be persisted in strict persistence scheme thus hurting the lifetime of NVM and incurring long writing latency.

Unlike strict persistence schemes, Osiris~\cite{YeHA18} recovers counter blocks with low writing overhead on execution time by relaxing counter block persistence and retrying counter multiple times during recovery process to obtain the correct one. However, Osiris only focuses on counter recovery and fails to recover the integrity tree. To recover the integrity tree, Triad-NVM~\cite{AwadYSNZ19} reconstructs the whole merkle tree from the bottom up and persists the lowest $N$ level nodes with data to trade off the number of memory writes with recovery time. However, Triad-NVM fails to be used in SIT, since the MAC in an SIT node can't be recalculated by the child nodes but relies on the correct counters in its own and the father node, which invalidates the bottom-up recovery approach. Furthermore, Anubis~\cite{ZubairA19} provides a fast recovery scheme for SIT. With each memory write (user data, counter block or SIT node), there is a metadata node changed in metadata cache, i.e., counter block or SIT node. Anubis uses a shadow table (ST) block to record the address of the changed metadata, the changed MAC and the Least Significant Bits (LSBs) of 8 counters in the changed node. The ST block is further written into NVM. Anubis could fast recover the SIT by only recovering the changed SIT nodes that are not flushed into persistent memory before system crashes according to the ST block, instead of recovering all the SIT nodes. However, Anubis incurs 2 times writes as described in~\cite{ZubairA19}, including normal memory writes and additional ST block writes, compared with traditional memory systems. More writes will hurt the lifetime of NVM and increase execution latency.

To fast recover the security metadata after system crashes with low write overhead, we propose STAR (\textbf{S}IT \textbf{t}race \textbf{a}nd \textbf{r}ecovery scheme) focusing on SIT, since its parallelizable MAC-computing provides good performance. STAR is flexible and adaptive to different configurations to obtain cost-efficient tradeoff between write overhead and recovery time, while delivering high performance, as shown in section~\ref{section 4.2}. STAR flushes the modified metadata node in cache by our proposed ahead write approach when a memory write causes changes in this metadata node. STAR also supports metadata to be restored and verifies the correctness of recovery process. To facilitate fast recovery, we need to address main three problems:

\begin{enumerate}
	\item{\emph{\textbf{The correctness of data.}}} Some security metadata in NVM are stale since the latest metadata in cache have not been flushed before crashes. Thus we need to restore these stale security metadata using the correct data on recovery. Different schemes have different approaches to obtain the correct data after system crashes. The strict persistence scheme doesn't need to obtain correct data since the security metadata have been already newest in NVM. Anubis stores the correct data in the shadow table block for SIT and flushes it with each memory write. Triad-NVM~\cite{AwadYSNZ19} obtains the upper-level MACs in a merkle tree by hashing the consistent lower-level MACs. cc-NVM~\cite{YangLCMS19} and Osiris~\cite{YeHA18} retry the counter multiple times and verify if the restored counter is correct or not according to ECC and MAC.
	\item{\emph{\textbf{The locations of stale data.}}}  Obtaining the correct data only ensures that the security metadata could be recovered after crashes, with the cost of a long recovery time if the system restores all security metadata. With the aid of the locations of stale security metadata, a system only needs to restore the stale nodes, significantly reducing the recovery time. In existing works, Anubis records the addresses of stale nodes in the ST block; Triad-NVM flushes $N$ low-level integrity tree nodes with user data and treats all upper-level nodes to be stale. Osiris doesn't record the locations of stale counter blocks. On recovery, Anubis only needs to restore the marked stale metadata, Triad-NVM needs to reconstruct the whole tree from the persistent low-level metadata, and Osiris needs to restore all counter blocks.
	\item{\emph{\textbf{The efficiency of verification.}}} Attackers can modify or replace a persistent memory line during recovery process. A verification mechanism is needed to ensure the correctness of the recovery. For a merkle tree, the root can be used to verify the correctness of recovery due to exhibiting each data change. But for the SIT lazy scheme~\cite{TaassoriSB18,SaileshwarNREJQ18} (described in Section~\ref{section 2.3}), SIT root can't verify the correctness of recovery due to failing to reflect the data changes immediately, and thus extra verification mechanism is needed to ensure the correctness of recovery process.
\end{enumerate}

To address the above problems, STAR stores the correct data, i.e., the correct counter of one node, in unused space of MAC in the child nodes (Section~\ref{section 4.3}). We observe that the memory writes from the user data cache have high spatial locality. Based on this insight, STAR places several bitmap lines in ADR to indicate which metadata line is stale in NVM (Section~\ref{section 4.4}). STAR constructs a cache-tree and uses the cache-tree root to verify the recovery process (Section~\ref{section 4.5}). To accelerate the recovery process, STAR introduces a multi-layer index structure to selectively read the useful bitmap lines in the NVM recovery area (Section~\ref{section 4.6}).

To evaluate the performance of our proposed scheme, we use Gem5~\cite{BinkertBBRSBHHKSSSSVHW11} with NVMain~\cite{PorembaZ015} to implement STAR, and run 8 benchmarks from SPEC2006 suite~\cite{Henning06} and 5 persistent workloads widely used in state-of-the-art works~\cite{CoburnCAGGJS11,RenZKCWM15,KolliRDSPLCW16,KolliGSDCNW17,LiuKRK18,Zuo018}. Our experimental results show that STAR incurs 1.13x/1.29x/2.01x memory writes of three configurations compared with traditional memory systems while one state-of-the-art work, Anubis, incurs 2x memory writes. With the different numbers of memory writes, STAR needs 0.039s/0.023s/0.004s recovery time for a 4MB metadata cache, while Anubis needs 0.020s. Moreover, the recovery time in STAR is only a function of the number of dirty nodes in metadata cache instead of memory and cache sizes, which offers the adaptability for the larger NVM and metadata cache. In summary, this paper makes the following contributions:
\begin{itemize}
	\item \textbf{A new metadata write approach.} We propose a new metadata cache write approach, called Ahead Write (AW). When a user data/metadata write causes its father node changed, the AW approach flushes the changed node with the user data/metadata and ensures the metadata consistency in NVM.
	\item \textbf{Counter-MAC synergization for restoring stale metadata.} We reuse the unused bits in the MAC in one node to store the correct metadata of its father node, which offers a new approach to restore stale metadata without any extra memory writes.
	\item \textbf{Bitmap lines for locating stale metadata.} To efficiently locate the stale metadata with low write overhead, we propose bitmap lines to absorb the location information. Using the bitmap lines to record location is useful for the applications with the high spatial locality.
	\item \textbf{Experimental evaluation.} We have implemented and evaluated STAR, and experimental results show STAR reduces the number of memory writes by up to 87\% with comparable recovery time compared with state-of-the-art work Anubis. STAR can also provide shorter recovery time with incurring negligible write overhead compared with Anubis. 
\end{itemize}

\section{Background and Motivation}
\label{section 2}
In this section, we present the background of recovering secure persistent memory and the motivation of our design.

\vspace{-0.2cm}
\subsection{Threat Model}
\label{section 2.1}
\vspace{-0.1cm}
In general, the processor chip is considered to be secure~\cite{YanEPRS06,RogersCPS07,AwadMHSH16,LiuKRK18,YeHA18,ZubairA19,AwadYSNZ19}. An attacker can attack the memory via multiple methods, such as scan the memory, snoop the memory bus and steal DIMM to obtain the user data, which exacerbate the data confidentiality. Attackers can also replay memory data and tamper with memory contents, which undermine the data integrity. Other attacks such as access pattern leakage, power analysis and side-channel attacks, are beyond the scope of this paper.

\begin{figure}[t]
	\vspace{-0.3cm}
	\centering
	\includegraphics[width=0.45\textwidth]{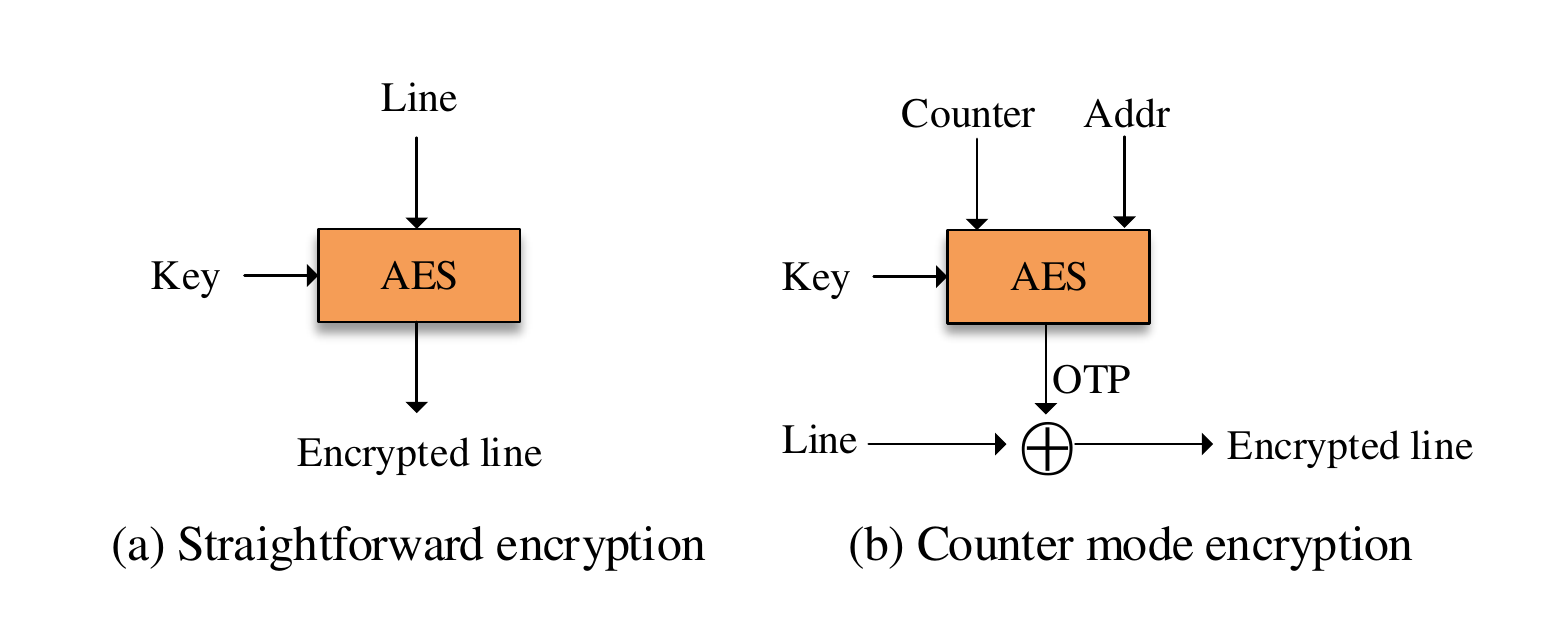}
	\vspace{-0.3cm}
	\caption{Different encryption schemes.}
	\label{AESandCME}
	\vspace{-0.45cm}
\end{figure}

\vspace{-0.2cm}
\subsection{Counter Mode Encryption}
\label{section 2.2}
\vspace{-0.1cm}
Since NVMs retain data after power failure, information leakage in NVMs is more severe than the volatile DRAM. User data encryption is necessary to protect data confidentiality. The data encryption can be processed in memory side~\cite{ChhabraS11} or processor side~\cite{YanEPRS06}. However, in the memory side encryption, the plaintext data should pass the memory bus and can be snooped by attackers. Thus state-of-the-art works use the encryption in the processor side~\cite{YoungNQ15,AwadMHSH16,SwamiRM16,Zuo0ZZG18}. A straightforward method in the processor side to encrypt a memory line is to use a block cipher algorithm, e.g., AES, with a global key, as shown in Fig.~\ref{AESandCME}(a). However, this straightforward encryption has some limitations. Due to the unchanged keys, attackers can easily break the encryption using dictionary attack. Besides, the decryption process is on the read critical path. The ciphertext data read from NVM need to be decrypted first, which introduces a long decryption latency.

Counter mode encryption (CME) is proposed to compensate for the above drawbacks of straightforward encryption. As shown in Fig.~\ref{AESandCME}(b), CME first uses counter, data line address and a global key to generate a one-time padding (OTP) via the AES algorithm. For memory writes, the cache line to be written needs to be encrypted by XORing the plaintext data with OTP. For memory reads, OTP is generated in parallel with reading memory line, and the plaintext data are obtained by XORing the line and OTP. Thus the decryption latency is hidden by the latency of reading data. To provide high security, OTP will be not reused. To meet this requirement, OTP generation uses three inputs, i.e., the line address, counter and key. Different data lines have different addresses, which allows the OTPs between different lines not to be reused. On the same line, each memory write causes its counter increased, thus OTP will not be reused in the same line at different writes. Usually, each counter block contains 64 7-bit minor counters and one 64-bit major counter. One counter block covers 64 user data blocks, i.e., one page. CME encrypts data blocks using corresponding minor counter and major counter. When one minor counter overflows, the major counter is increased by one. All the minor counters are reset and the data blocks in this page need to be re-encrypted. The 64-bit major counter never overflows throughout the lifespan of an NVM since the count range, i.e., $2^{64}$ $\approx$ $10^{20}$, is far larger than the endurance limit of NVM cell, e.g., $10^{7}$-$10^{9}$ for PCM~\cite{QureshiKFSLA09,ZhouZYZ09} and $10^{8}$-$10^{12}$ for ReRAM~\cite{lee2010evidence,lee2011fast}.

\begin{figure}[t]
	\vspace{-0.3cm}
	\centering
	\includegraphics[width=0.45\textwidth]{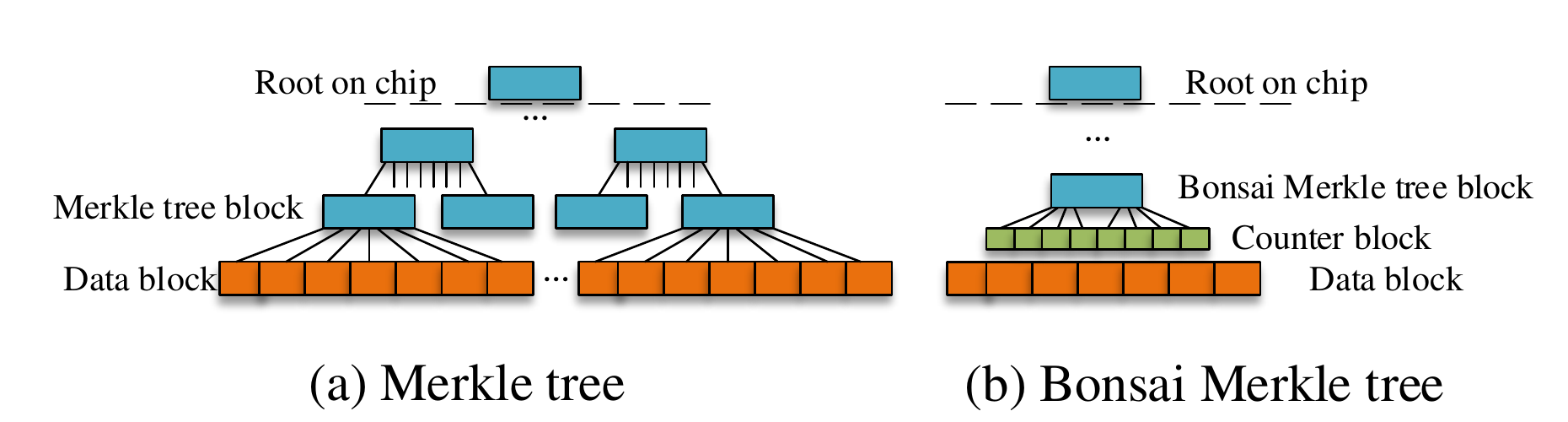}
	\vspace{-0.3cm}
	\caption{Different bottom-up integrity trees. (a) Merkle tree starts hashing from data blocks; (b) Bonsai Merkle tree starts hashing from counter blocks.}
	\label{integrity_tree}
	\vspace{-0.45cm}
\end{figure}

\begin{figure}[t]
	\vspace{-0.3cm}
	\centering
	\includegraphics[width=0.45\textwidth]{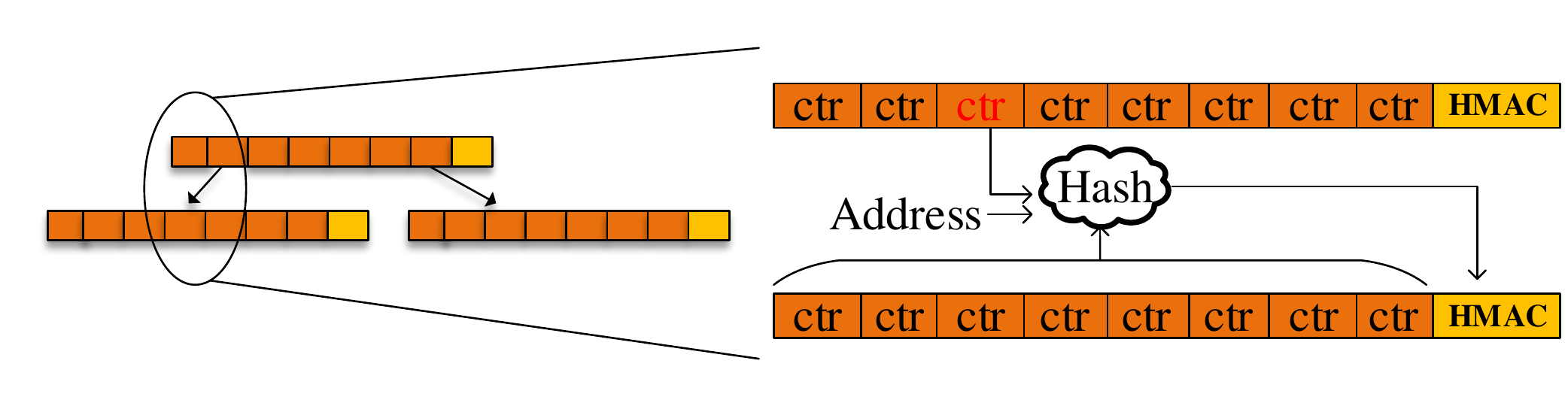}
	\vspace{-0.3cm}
	\caption{SGX integrity tree (SIT) that generating the MAC by hashing node address, all counters in this node and one corresponding counter in the father node.}
	\label{SIT}
	\vspace{-0.45cm}
\end{figure}

\vspace{-0.2cm}
\subsection{Integrity Tree}
\label{section 2.3}
\vspace{-0.1cm}
Integrity trees are widely used in memory systems to protect data integrity, e.g., merkle tree, bonsai merkle tree (BMT) and SGX integrity tree (SIT)~\cite{GassendSCDD03,RogersCPS07,CostanD16,TaassoriSB18}. As shown in Fig.~\ref{integrity_tree}(a), several data blocks are hashed together to generate MAC using a cryptographic hash function stored in the processor. Higher level MAC is generated by hashing lower level MACs together and finally, the merkle tree is formed with one 64B hash root stored on chip. Bonsai merkle tree generates the first MAC level by hashing counter blocks instead of user data as shown in Fig.~\ref{integrity_tree}(b). Since the number of counter blocks is far smaller than that of user data blocks, bonsai merkle tree has fewer leaf nodes and lower heights than merkle trees. Since each MAC in one node needs its child node MACs as inputs, the (bonsai) merkle trees calculate their MACs serially, i.e., the tree can't calculate one MAC before completing the computing upon the MACs of its child nodes.

Different from merkle tree and bonsai merkle tree, SIT node contains 8 56-bit counters and one 64-bit MAC~\cite{TaassoriSB18} instead of hash values. As shown in Fig.~\ref{SIT}, the MAC in each SIT node is generated by hashing the node address, the counters in this node and one corresponding counter in the father node. SIT can calculate different level MACs in parallel as long as these counters have been increased correctly.

When a user data block is written from cache to memory, (bonsai) merkle tree needs to change all nodes in the branch from the data block to the root and finally, the root is modified to show the changes of data. Unlike (bonsai) merkle tree, SIT has two options to update the tree nodes, including lazy and eager schemes. In the lazy scheme, once a data block, i.e., user block, counter block or SIT node, is flushed into NVM, its ancestor SIT nodes up to the first cache hit are cached to verify the integrity of the flushed data block. Then the corresponding counter in the father node is increased, and the MAC in the flushed block is modified due to the increased corresponding counter. The ancestor nodes including the root are unchanged, and they are modified only when their dirty child nodes in cache gets evicted later. Thus the SIT root is not modified immediately with the changes of data and intermediate nodes in the lazy scheme. Moreover, the eager scheme is to propagate the changes to the root and modify root immediately when one data block or SIT node is evicted into NVM. The root shows the data changes and has higher probability of overflowing since each data write, no matter where it is written, causes a counter to be increased in the root. In this paper, we use the lazy approach to update the SIT as Synergy~\cite{SaileshwarNREQ18}, Vault~\cite{TaassoriSB18} and Anubis~\cite{ZubairA19}.

\vspace{-0.2cm}
\subsection{The MAC in Persistent Memory}
\label{section 2.4}
\vspace{-0.1cm}
SIT uses MAC stored in a node to associate counters in this tree node with one counter in its father node. Any unauthorized modifications over these counters and MAC could be detected. The user data also needs MAC to associate the data with encryption counters by hashing the data, block address and corresponding counter to generate the MAC. Without the MAC, the illegal modification in the user data can't be detected and the wrong data will be used by CPU after decryption. When reading, the integrity of user data needs to be verified using MAC. To avoid the failure of integrity checking on recovery, MAC needs to be written into NVM with the new user data. MAC is logically placed with the user data line, but physically placed in another MAC memory line. To reduce the MAC memory line access, especially memory line writing, Synergy~\cite{SaileshwarNREQ18} stores the MAC in the 9th chip, in which ECC is previously stored. Synergy reads/writes the data and MAC in one memory access. In this paper, we use the result of Synergy to store the data and MAC in one line instead of two memory lines.

In general, the size of MAC is 64 bits~\cite{TaassoriSB18}. However, 54-bit MAC is also secure as described in Morphable Counters~\cite{SaileshwarNREJQ18}. There are 10 unused bits in a 64-bit MAC space now. We will reuse these unused bits in our design.

\vspace{-0.2cm}
\subsection{Motivation}
\label{section 3}
\vspace{-0.1cm}
The state-of-the-art works provide the metadata recovery schemes in the persistent memory after system crashes, e.g., Osiris~\cite{YeHA18} and Triad-NVM~\cite{AwadYSNZ19}, which fail to support SIT recovery, and Anubis~\cite{ZubairA19}, which needs high write overhead. Osiris relaxes the counter block persistence during system running time. After system crashes, Osiris recovers the stale counter block by retrying the counter from the stale counter to stale counter+N, checks the correctness of this counter recovery by ECC and finally detects the data replace attack by using merkle tree root stored on chip. However, on SIT, the root on chip doesn't reflect the latest data in memory. Attackers can simply replace the data, MAC and ECC with old tuple on recovery and this data replacement can't be verified when Osiris is used for SIT. Osiris also needs a long time to recover the counter blocks due to failing to distinguish the stale counter blocks and restoring all counters. Triad-NVM flushes multiple-level merkle tree nodes with the user data into NVM. The merkle tree recovery on Triad-NVM needs to reconstruct the whole merkle tree and compare the reconstructed root with the root stored on chip. But SIT can't be constructed from the leaves. Without the correct corresponding counter in the father node, the MAC in one node even cannot be computed. For Triad-NVM, although the low-level nodes in the merkle tree are unnecessary to be recovered, reconstructing the whole tree has to consume a long time.

Anubis provides fast recovery schemes for both merkle tree and SGX integrity tree. For a merkle tree, when a metadata block in cache is marked dirty from a clean state, an extra block with this dirty block address is written into NVM. For the SGX integrity tree, each metadata write from cache into memory will incur an extra block write containing the address, the LSB of counters and MAC of the father node of the written metadata block. The recovery scheme in Anubis for SIT has 2 times writes compared with normal write-back scheme, increasing the application execution time and reducing the lifetime of NVM.

\begin{figure}[t]
	\vspace{-0.3cm}
	\centering
	\includegraphics[width=0.45\textwidth]{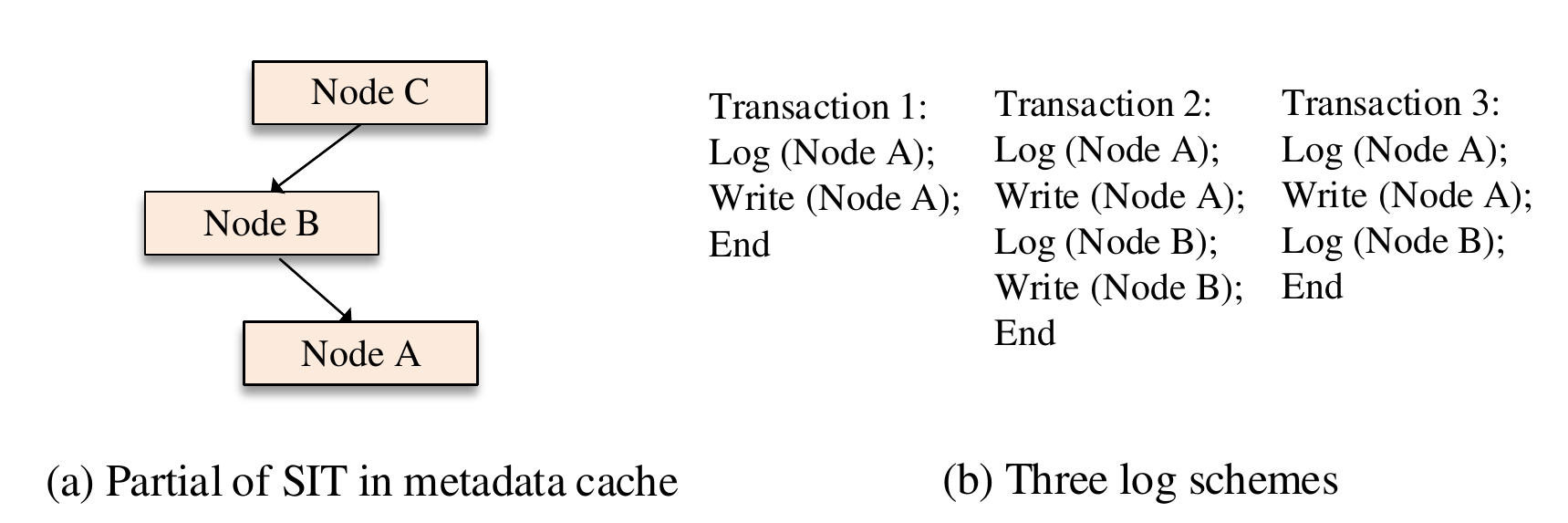}
	%\vspace{-0.3cm}
	\caption{Using log schemes to persist the SIT nodes.}
	\label{log}
	\vspace{-0.45cm}
\end{figure}

A log is usually used to provide the data consistency, e.g., redo log and undo log. For redo log, the new data is first written into the log and then the old data is updated in-place. If a system crash occurs during writing log, the old data in-place is consistent; if the crash occurs during updating old data, the inconsistent old data can be recovered according to the new data recorded in the log. An undo log is also used to recover the inconsistent data by the old data in a log. However, SIT cannot use logs to ensure the consistency between two SIT nodes. As shown in Fig.~\ref{log}(a), node B is the father node of node A and node C is the father node of node B. The metadata cache needs to evict the node A due to the cache replacement policy. Fig.~\ref{log}(b) shows three different log schemes for persisting node A. Transaction 1 ensures the node A itself in NVM is consistent. However, the node B has been modified due to the eviction of its child node A. After system crashes, node B in NVM is inconsistent to its copy in cache and needs to be restored. Transaction 2 provides the consistency of both nodes A and B, but the node C is inconsistent if system crashes occurred. Transaction 3 writes the node A and logs the node B. When a crash occurred, although the node B is inconsistent, it could be restored from the log. This transaction provides the consistency of nodes A and B. This log scheme is similar with Anubis, i.e., when writing node A, Anubis uses the ST node to record and restore the modified node B while Transaction 3 uses the log.

In this paper, we focus on SIT and counter mode encryption to propose a new scheme, i.e., STAR, to reduce the extra block writes on running time with fast recovery after system crashes.

\section{System Design}
\label{section 4}
In this section, we describe the design details of STAR in the context of the SGX integrity tree in NVM.

\begin{figure}[t]
	
	\centering
	\includegraphics[width=0.45\textwidth]{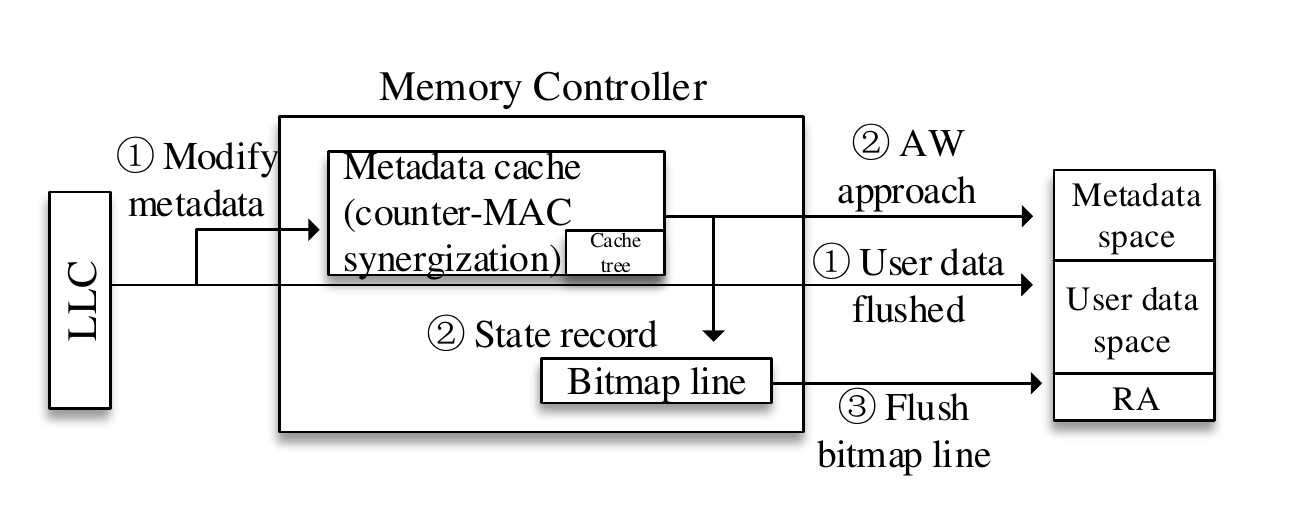}
	\vspace{-0.3cm}
	\caption{The overview of STAR design. \textcircled{1} Flushing user data incurs the modifications of counters in metadata cache. \textcircled{2} When the metadata with counter-MAC synergization are flushed by Write Ahead approach, their dirty/clean states are recorded in bitmap lines. \textcircled{3} Bitmap lines are flushed into recovery area by LRU policy. Cache-tree exhibits each change in metadata cache.}
	\label{overview}
	\vspace{-0.45cm}
\end{figure}

\vspace{-0.2cm}
\subsection{STAR Overview}
\label{section 4.1}
\vspace{-0.1cm}
A naive approach to recover the security metadata after crashes is to use strict persistence scheme that persists all the nodes in a branch from counter block to the root. This scheme needn't to restore metadata after crashes since the metadata has been consistent in NVM. However, the strict persistence scheme incurs many memory writes, hurting the lifetime of NVM and performance during running time. In the context of SIT, our STAR provides configurations to meet different requirements on write overhead and recovery time. As shown in Fig.~\ref{overview}, STAR consists of Ahead Write approach (AW), counter-MAC synergization in metadata cache, bitmap lines and cache-tree. Any data flushing from the last level cache (or metadata cache) into NVM causes the modifications of its father node in metadata cache (\circled{1}). Counter-MAC synergization leverages the unused bits in MAC space of one node to store the LSBs of modified counter in its father node. On recovery, the stale node could be restored from its child node. When metadata nodes (counter blocks and SIT nodes) are evicted from metadata cache, the AW approach is used to flush the metadata and the state of metadata node is recorded in bitmap lines (\circled{2}).The AW approach provides the consistency between metadata in cache and their copies in NVM by ahead writing the modified father node of the flushed metadata. Although the AW approach is used, the stale nodes also need to be recovered, and the correctness of recovery process should be verified, i.e., whether an attack occurs during the recovery process. The bitmap lines are used to record the locations of nodes whose states changed, especially indicating which node is dirty, to provide fast recovery feature of STAR. Bitmap lines are stored in an ADR region in the memory controller and flushed into the recovery area in NVM by LRU policy (\circled{3}). Finally, the cache-tree in metadata cache shows the changes in each location in the metadata cache. When the last level cache or metadata cache writes incur the changes in metadata cache, e.g., the metadata modification and eviction, the cache-tree and its root will be modified to show the metadata changes. During recovery, the cache-tree with bitmap lines is used to detect whether an attack occurs during the recovery process.

We further introduce the main components of STAR.

\begin{table}[b]
	%\vspace{5px}
	\footnotesize
	%\scriptsize
	%\tiny
	\vspace{-0.2cm}
	\caption{\label{table:trade-off} The different configurations of STAR provide the different write and recovery overheads.}
	\vspace{-5px}
	\begin{center}
		\begin{tabular}{|c|c|c|}
			\hline
			\textbf{AW start-level} & \textbf{Write overhead} & \textbf{Recovery overhead}\\
			\hline
			User data (AW-L) & High & \tabincell{c} {Not recover counter \\ blocks and SIT nodes}\\
			\hline
			Counter block (AW-M) & Normal & \tabincell{c} {Recover only counter blocks}\\
			\hline
			(N-1)th level (AW-H) & Low& \tabincell{c} {Recover counter \\ blocks and SIT nodes}\\
			\hline
		\end{tabular}
	\end{center}
	%	\vspace{-7px}
\end{table}

\begin{figure}[t]
	\vspace{-0.3cm}
	\centering
	\includegraphics[width=0.45\textwidth]{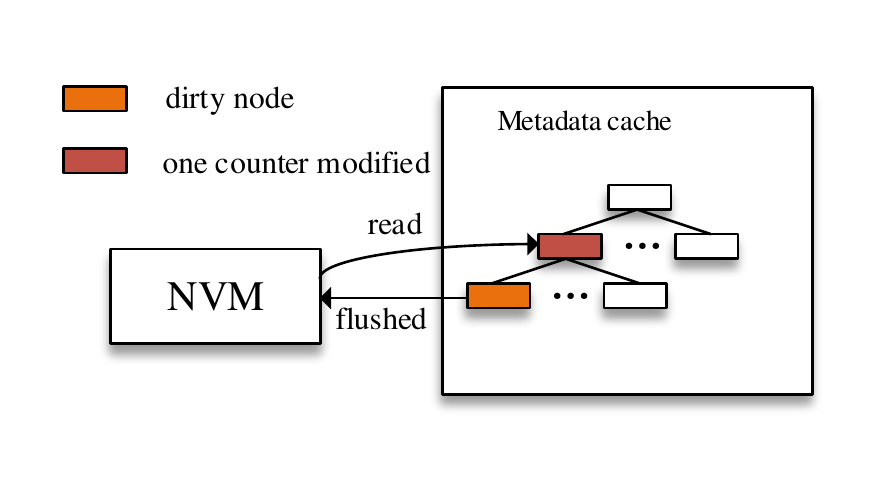}
	\vspace{-0.6cm}
	\caption{An SIT node becomes/stays dirty when its dirty child node is evicted from metadata cache.}
	\label{node_become_dirty}
	\vspace{-0.45cm}
\end{figure}

\begin{figure}[t]
	\vspace{-0.3cm}
	\centering
	\includegraphics[width=0.45\textwidth]{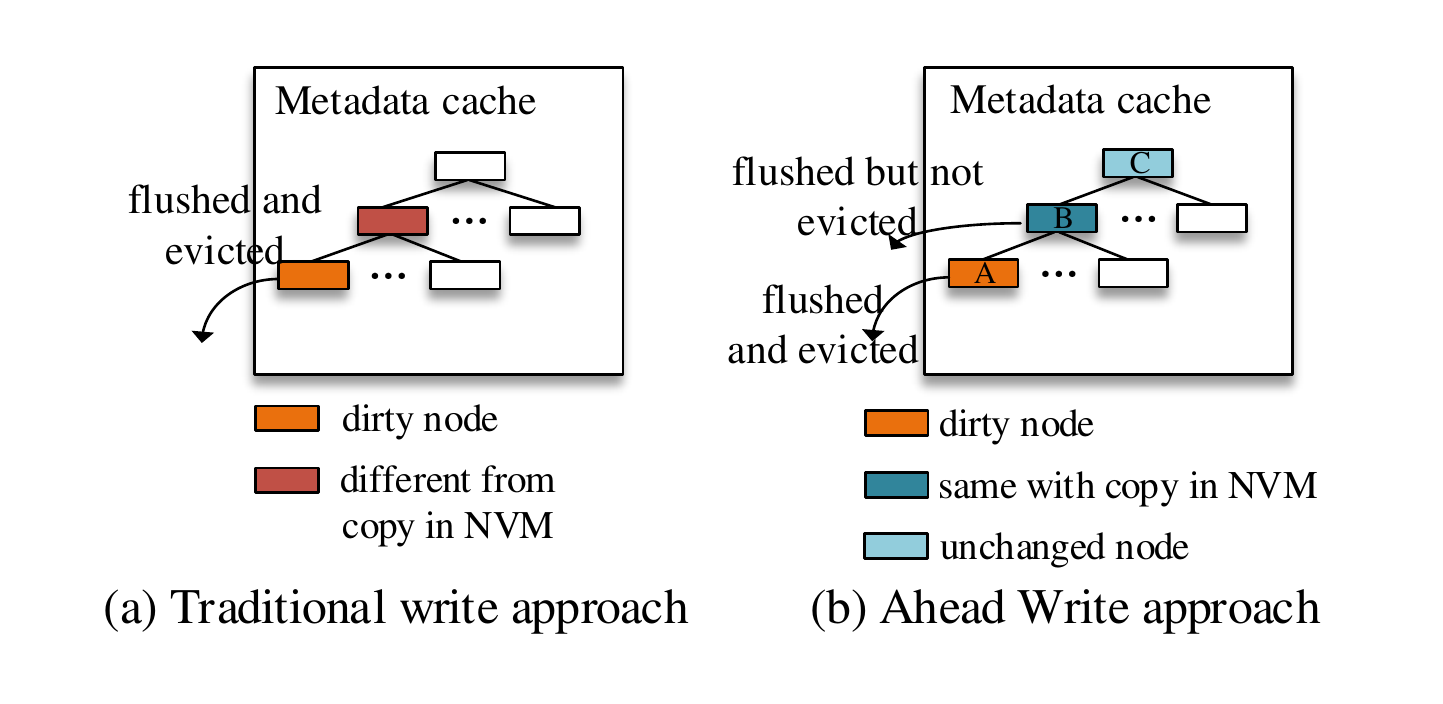}
	\vspace{-0.3cm}
	\caption{Different write approaches to flush the SIT node. (a) Traditional write approach only flushes the evicted node; (b) The Ahead Write approach flushes the evicted node and its father node, but still keeps the father node in the metadata cache.}
	\label{Ahead_write}
	\vspace{-0.6cm}
\end{figure}

\vspace{-0.2cm}
\subsection{Ahead Write Approach}
\label{section 4.2}
\vspace{-0.1cm}
As shown in Fig.~\ref{node_become_dirty}, when a dirty metadata node is evicted from cache, if its father node is not in cache, the father node and ancestor nodes need to be cached until the first metadata cache hit to verify the integrity of the evicted node. Then in the father node, the corresponding counter of the flushed one is increased by one. We observe that an SIT node will be modified when its \textbf{dirty} child node is \textbf{evicted} from metadata cache. If a clean node has been evicted from cache, its father node will not be modified. Moreover, if a node is changed dirty from clean state in cache, its father node will not be modified. Based on the above observation, we propose Ahead Write (AW) approach to ensure the consistency of metadata nodes in cache and their copies in NVM. Note that if a cached metadata is consistent with its copy in NVM, the metadata needn't to be restored after crashes. As shown in Fig.~\ref{Ahead_write}(a), a memory system only flushes the evicted metadata node itself, and the father node in cache is modified making the father node different from its copy in NVM. In our AW approach in Fig.~\ref{Ahead_write}(b), we flush not only the evicted node itself but also its changed father node. The AW approach updates the father node in NVM, but doesn't evict it from cache. It means that the father node is still in cache and all ancestor nodes are not modified according to our observation above. Flushing the evicted node and its changed father node ensures that all the changes occurring in metadata cache have been flushed into NVM. However, the AW approach incurs extra 1x write compared with traditional write approach, i.e., one write for evicted node and one write for the changed father node. In fact, AW approach can reduce most extra writes for metadata nodes since the metadata node to be evicted is latest in NVM. As shown in Fig.~\ref{Ahead_write}(b), node A is evicted from metadata cache before its father node B. When node A is evicted, AW approach \textbf{flushes and evicts} the node A and \textbf{flushes but doesn't evict} its father node B into NVM (node B is still in cache). Later, when the node B is evicted from cache, since the latest node B exists in NVM, the AW approach needn't to flush the node B into NVM, but discards it from cache. Instead, the AW approach writes the father node C, which is modified due to the eviction of node B, into NVM and still keeps the node C in cache, meaning that node C is identical to its copy in NVM. If the node B is evicted before its child node A, the node B has not been modified since its child node is still in cache. The AW approach discards but doesn't flush the node B like traditional approaches. When evicting a metadata node from cache, AW approach only changes the state of evicted node to the clean like a normal write, The state of the modified father node which is written in advance from cache is still dirty.

In summary, the AW approach incurs two writes on one SIT node (called start-node) and one write on the ancestor nodes. Our AW approach manages these start-node in the same level which we call start-level and incurs one write on the upper levels. The SIT nodes higher than start-level are ensured to be flushed into NVM as long as they are modified in cache. The SIT nodes in start-level and lower levels than start-level need to be recovered after crashes since their copies in NVM are stale. To meet the requirements of different write and recovery overheads, STAR tries different AW start- levels. As shown in Table~\ref{table:trade-off}, in terms of recovery time, when the user data is AW start-level (we call this configuration as AW-L), all the security metadata nodes, i.e., counter blocks and SIT nodes, are higher than the AW start-level. Thus the metadata nodes need not to be restored after crashes. When the counter block is AW start-level (we call this configuration as AW-M), only the counter blocks need to be restored. Moreover, if the (N-1)th level of SIT, whose father-level is root, is AW start-level (we call this configuration as AW-H), all the security metadata nodes need to be restored. In terms of write overhead, AW-L introduces high write overhead since every user data write will incur its father node write, i.e., writing the counter block. Since the frequency of flushing counter blocks is lower than that of user data, the write overhead introduced by AW-M is lower than AW-L. AW-H doesn't incur extra write overhead since its father node is root on-chip, which is never written into NVM.

We analyze the security of the AW approach. As shown in Fig.~\ref{Ahead_write}(b), the node B is ahead written into NVM when the child node A is evicted from metadata cache. Since the node C is untouched, attackers can replace the new node B with the old one and the untouched node C matches both the new and old node B. Fortunately, during running time, if the latest node B is in cache, the system will use the correct cached one instead of reading the replaced node B from NVM. If the node B is evicted from cache, the corresponding counter in the father node C is increased by one and the replace attack could be detected since the replaced node B in can't match its father node C in cache. During running time, the AW approach is secure. However, if the replace attack occurs during the recovery process, replacing the node B in NVM with the old one can't be detected since the latest node B is not in cache and its father node C has not been modified. Our STAR needs a mechanism to detect the replace attack during the recovery process. Moreover, AW-M and AW-H also need to restore the metadata as shown in Table~\ref{table:trade-off}.

We further present the counter-MAC synergization, bitmap lines and cache-tree to restore the stale metadata and verify the correctness of recovery process.

\begin{figure}[t]
	\vspace{-0.3cm}
	\centering
	\includegraphics[width=0.50\textwidth]{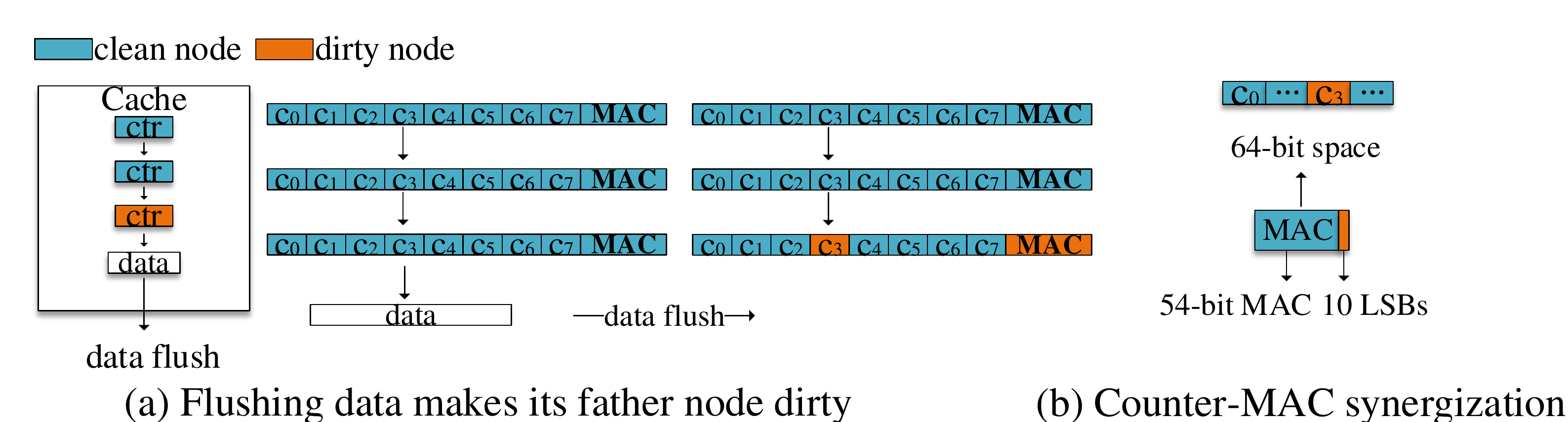}
	\vspace{-0.3cm}
	\caption{A new approach to restore dirty nodes. (a) Once one user data is written, only its corresponding counter in the father node is dirty; (b) 10 Least Significant Bits of dirty father counter is stored in the MAC space of child node.}
	\label{dirty_node}
	\vspace{-0.45cm}
\end{figure}

%\vspace{-0.2cm}
\subsection{Restoring the Metadata}
\label{section 4.3}
\vspace{-0.1cm}
The metadata in cache and NVM could be divided into two types, i.e., clean and dirty nodes. The clean nodes in cache are the same as their copies in NVM, or the clean nodes have not been cached only in NVM. The dirty nodes in cache are different from their copies in NVM since the cached one has been modified but not been flushed into NVM. In other words, the clean nodes in NVM are latest and the dirty nodes are stale. A cache has a ‘dirty bit’ for each cache line to indicate if this cache line is consistent with its copy in NVM or not. We distinguish the clean and dirty nodes by checking the dirty bit.

On recovery, we only need to restore the stale security metadata nodes in NVM. One key observation is that in SIT lazy update scheme (detailed in Section~\ref{section 2.3}), when data is evicted from cache, only the corresponding counter in the father node of the evicted data is increased by one. The MAC in the father node is modified, and other nodes are untouched as shown in Fig.~\ref{dirty_node}(a). After obtaining the correct copy of the modified counter, we re-compute the correct MAC by hashing the counters in this node and one corresponding counter in the father node. Moreover, a MAC needs to be persisted with data to associate the data with its father node. For example, if the latest user data MAC has not been persisted, the data can't be used after recovery since the stale MAC can't match the latest data and counters stored in NVM. As described in Section~\ref{section 2.4}, 54-bit MAC is also secure~\cite{SaileshwarNREJQ18} and 10 bits are unused in the 64-bit MAC space. STAR leverages these unused bits of MAC in the child node to store 10 LSBs of the corresponding counter in the father node, as shown in Fig.~\ref{dirty_node}(b). When a dirty node is to be written, its corresponding counter in the father node increases by 1. STAR stores the 10 LSBs of the modified counter in the unused space of the MAC to be written, which is called counter-MAC synergization. The 10 LSBs of a modified counter is flushed with the child node as long as the child node eviction causes the modification of counter in the father node. To protect the LSBs, MAC in a node is computed by hashing this node, the address of the node, the corresponding counter in the father node and the LSBs stored in the MAC space. When one counter in a node has been increased $2^{10}$ times, the node needs to be flushed into NVM to update the 44 Most Significant Bits (MSBs). This counter overflow is rare and introduces negligible overheads.

STAR obtains the correct LSBs of counter in the dirty node from its child nodes. Combined with the MSBs stored in the corresponding counter in the stale node, the stale counter is correctly restored and the MAC in the node could be re-computed. However, like the AW approach, restoring node according to counter-MAC synergization suffers from data replace attack. For example, when recovering a counter block of one user data, attackers could replace the user data, MAC and LSBs with old tuple and the modifications can't be detected as described in Section~\ref{section 4.5}. The counter-MAC synergization also requires a mechanism to verify the correctness of recovery process as the AW approach.

\begin{figure}[t]
	\vspace{-0.3cm}
	\centering
	\includegraphics[width=0.45\textwidth]{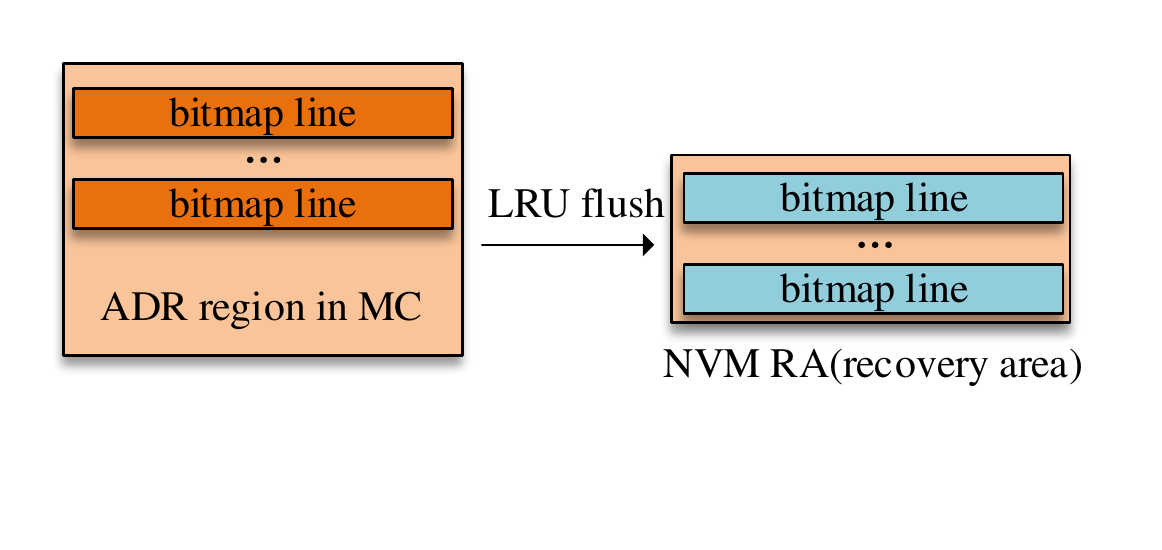}
	\vspace{-0.9cm}
	\caption{Bitmap lines are placed in ADR in Memory Controller and flushed into NVM recovery area (RA) by LRU. Each bitmap line covers 32KB metadata space.}
	\label{bitmap_line}
	\vspace{-0.45cm}
\end{figure}

\vspace{-0.2cm}
\subsection{Tracking the Stale Node Location}
\label{section 4.4}
\vspace{-0.1cm}
The locations of stale nodes in NVM are necessary for fast recovery. Without the locations, all the metadata nodes in NVM need to be restored due to staleness. To efficiently record the locations of the stale metadata nodes, bitmap lines are presented in Fig.~\ref{bitmap_line}. One bit in the bitmap line represents a security metadata line, e.g., the first and last bits in the first bitmap line represent the 1st and 512th metadata lines in metadata space. Each bitmap line covers 32KB continuous metadata space since one memory line contains 512 bits (512$\times$64B=32KB).

To ensure the persistence of bitmap lines, STAR leverages the asynchronous DRAM refresh (ADR) mechanism. Modern processor vendors provide a small battery backup for ADR with tens of entries~\cite{LiuKRK18,ShinTTS17} in memory controller. When the system crash occurs, the data stored in ADR could be flushed into NVM by battery backup support. STAR places a certain number of the bitmap lines in ADR (default is 16 lines). When a metadata line becomes dirty from the clean state according to its 'dirty bit', the corresponding bitmap line bit is set to 1, If a dirty metadata line is written from cache making it become clean from the dirty state, the corresponding bitmap line bit will be reset. If the bitmap lines in ADR don't cover one metadata line, whose corresponding bit is in another bitmap line, STAR flushes one bitmap line from ADR to the Recovery Area (RA) in NVM by LRU policy and creates a new bitmap line to record the state of metadata line. Note that when creating a new bitmap line, all the metadata cache lines covered by the new bitmap line should be traversed to identify the states. These states need to be recorded in the corresponding bits in the new bitmap line. Thus STAR never reads bitmap lines from RA during running time. RA in NVM is allocated to store all the bitmap lines, and consumes a negligible 1/512 space of the metadata space.

For AW-L and AW-M, although the cached nodes in the levels higher than the start-level are the same as their copies in NVM, STAR also sets/resets the corresponding bits to 1/0 in bitmap lines according to the 'dirty bit' when their child nodes/these nodes are flushed. This address information is used to verify the correctness of recovery process (detailed in Section~\ref{section 4.7}). 

\iffalse
\begin{figure}[t]
	\vspace{-0.3cm}
	\centering
	\includegraphics[width=0.50\textwidth]{design/cache_tree.pdf}
	\vspace{-0.6cm}
	\caption{Cache-tree. (a) Use dirty way in set to construct Set-MAC. (b) Use set-MACs to construct a cache-tree.}
	\label{cache_tree}
	\vspace{-0.45cm}
\end{figure}
\fi

\begin{figure}[t] \centering
	\vspace{-0.3cm}
	\subfigure[Set-MAC]{
		
		\begin{minipage}[b]{0.45\textwidth}
			
			%\label{cache_tree(a)}
			\includegraphics[width=1\textwidth]{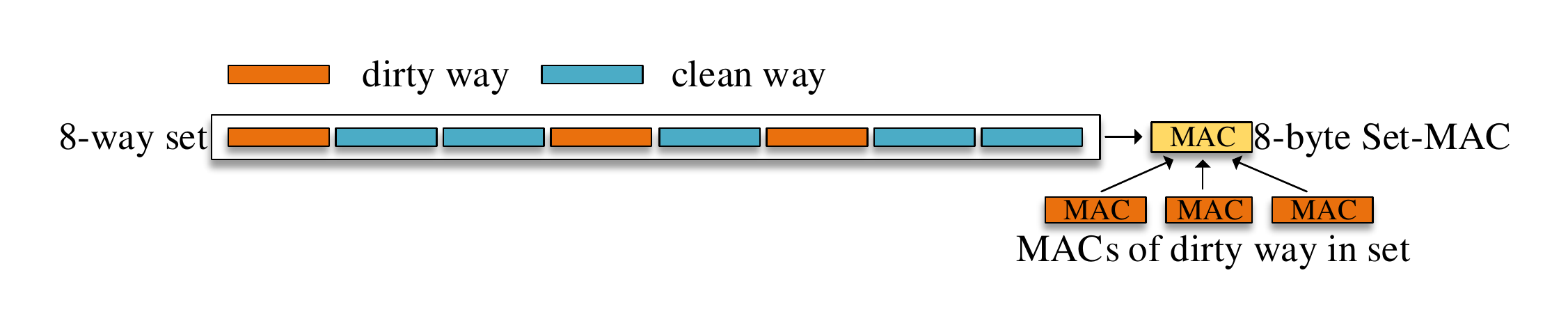}
			\vspace{-0.6cm}
		\end{minipage}
		
	}
	\vspace{-0.3cm}
	
	\subfigure[Cache-tree]{
		\begin{minipage}[b]{0.45\textwidth}
			
			%\label{cache_tree(b)}
			\includegraphics[width=1\textwidth]{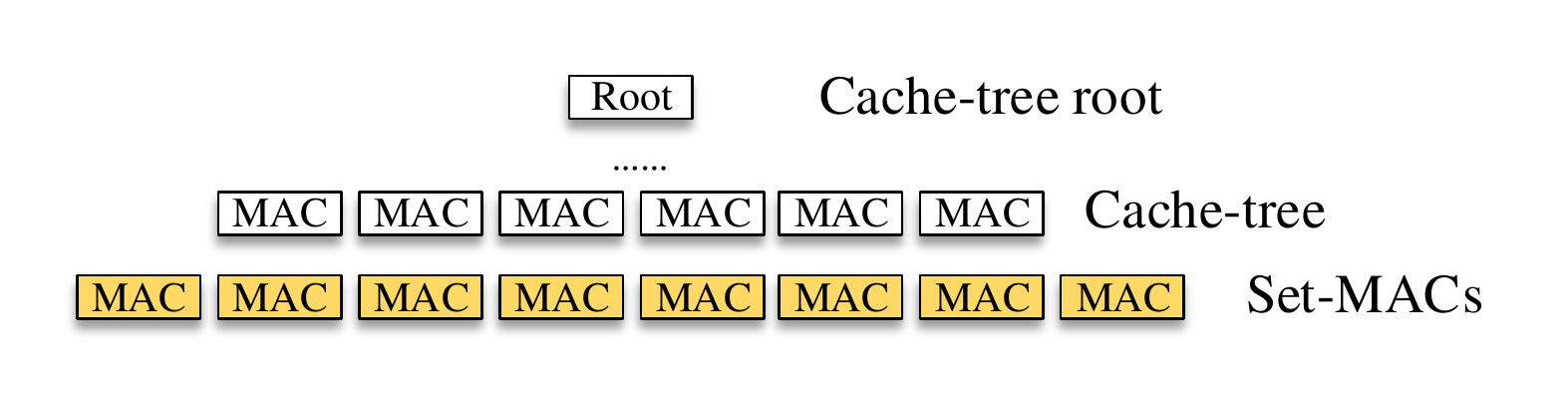}
			\vspace{-0.6cm}
		\end{minipage}
		
	}
	\vspace{-0.1cm}
	\caption{Using MACs in dirty metadata cache line to construct a cache-tree.}
	\label{cache_tree}
	\vspace{-0.6cm}
\end{figure}

\vspace{-0.1cm}
\subsection{Using Cache-Tree to Verify Recovery Process}
\label{section 4.5}
\vspace{-0.1cm}
STAR restores stale metadata nodes according to the counter-MAC synergization. However, attackers can replace the data, MAC and LSBs with an old tuple to disable the recovery without a system detection. For example, a user data with the LSBs 0x11 and MAC is written into NVM. During recovery, restoring its father counter block needs the LSB 0x11, but attackers replace the tuple with old data, old MAC and old LSBs 0x10. The replace attack can't be detected since the MAC in the user data matches the old tuple of data and LSBs and a wrong MAC is re-computed in the counter block. To detect this attack, we need a mechanism to verify if the recovery process is correct or not. 

We notice that modern cache usually is a set-way structure. An 8-way cache is divided into multiple sets. Each set has 8 ways and each way contains one cache line. A specific memory line is cached into a specific set and placed in any way in this set. As shown in Fig.~\ref{cache_tree}, STAR constructs a cache-tree by using the dirty node MACs to verify the correctness of the recovery process. MACs in dirty nodes in one set are first ordered by the descending addresses. Then the set-MAC is computed by hashing these ordered MACs of the dirty nodes as shown in Fig.~\ref{cache_tree}(a). Finally, a merkle tree is constructed by iteratively hashing these set-MACs as shown Fig.~\ref{cache_tree}(b) (called cache-tree). The cache-tree needs to be updated when a modification/eviction of dirty node occurs. If no dirty metadata lines are in one set, STAR uses zero -bytes as the set-MAC to construct the cache-tree. Note that STAR logically constructs the cache-tree without moving any cache line. STAR doesn't add an 8-byte space at each set. The set-MACs and cache-tree nodes existing in metadata cache with SIT nodes and counter blocks, and can also be evicted from cache. Note that the cache-tree nodes in cache are not involved in the set-MAC generation. The cache-tree root is always on chip just like a traditional merkle tree root. On recovery, STAR reconstructs the cache-tree. The replace attack will cause a wrong data MAC and be detected by the cache-tree root as shown in Section~\ref{section 4.7}.

\begin{figure}[t]
	\vspace{-0.3cm}
	\centering
	\includegraphics[width=0.45\textwidth]{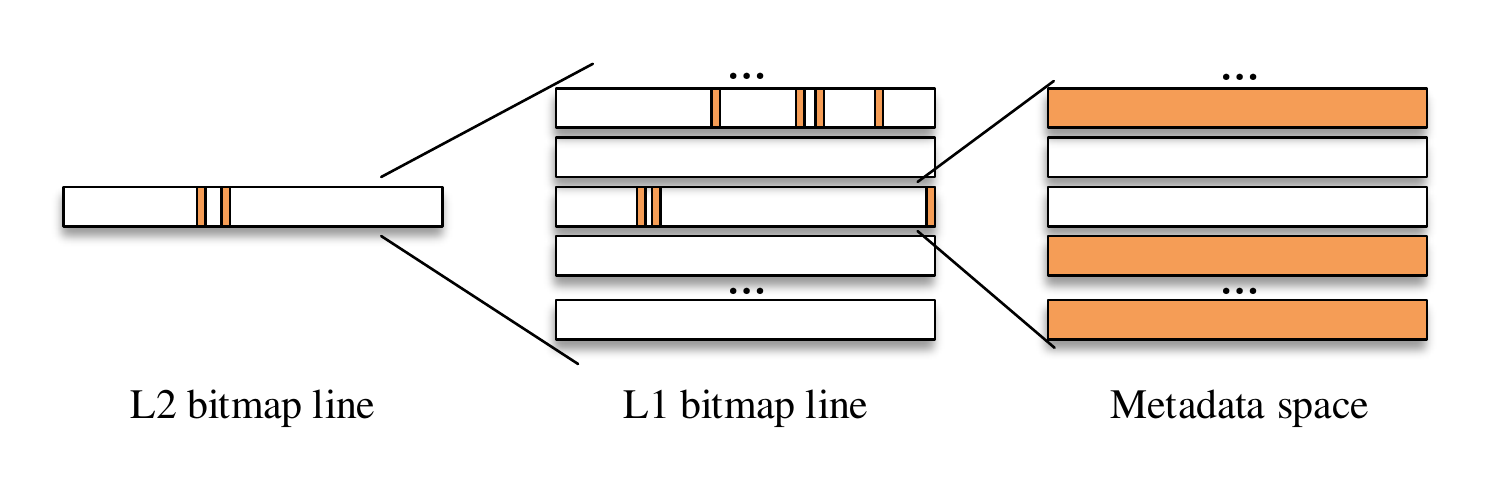}
	\vspace{-0.3cm}
	\caption{A multi-layer index structure is used to read the non-zero bitmap lines and stale metadata nodes.}
	\label{index}
	\vspace{-0.45cm}
\end{figure}

\vspace{-0.2cm}
\subsection{Using Multi-layer Index to Speed Up Recovery}
\label{section 4.6}
\vspace{-0.1cm}
To locate the stale metadata node in NVM, STAR needs to read all bitmap lines in RA. For a 16-GB NVM, the RA occupies 1/512 space of the metadata space, i.e., RA occupies a 4-MB NVM space. However, reading the 4-MB RA needs long latency compared with Anubis that only needs to read Shadow Table whose size is the same as that of metadata cache. To speed up STAR recovery process, we observe that only the locations of the stale nodes are needed. Even all the metadata cache lines are stale, a larger part of the bitmap lines in RA are useless during recovery, since bitmap lines are designed to cover all the metadata space, which is much bigger than the metadata cache. Reading the zero lines which don't record the locations of stale node from RA is useless and time-consuming. It is better to only read the non-zero bitmap lines instead of reading all lines from RA.

To speed up reading RA, we propose a multi-layer index, as shown in Fig.~\ref{index}. STAR leverages L1 bitmap lines to indicate which security metadata lines are stale and L2 bitmap lines to indicate which L1 bitmap lines are non-zero. If necessary, STAR can add L3 bitmap lines and so on. We call this structure multi-layer index. STAR stores the highest-layer bitmap lines on chip as SIT root and never flushes it into NVM. To reduce the consuming of on-chip space, the number of bitmap lines in the highest layer is always one. Other-layer lines are in ADR in the memory controller and flushed into RA by LRU like the bitmap lines described in Section~\ref{section 4.4}, consuming the negligible NVM space. A 1-/2-/3-layer index can cover 32KB/16MB/8GB metadata space. In our evaluation, we model 16GB main memory (about 2GB metadata) and the 3-layer index is sufficient.

\vspace{-0.2cm}
\subsection{Recovery Process}
\label{section 4.7}
\vspace{-0.1cm}
After crashes, the security metadata need to be recovered to the latest state. To recover the stale security metadata, STAR first reads the multi-layer index from RA to obtain the non-zero L1 bitmap lines. According to the L1 bitmap lines, STAR distinguishes the stale metadata node in metadata space. For different configurations of the AW approach, the number of nodes that need to be restored is different. For example, AW-H needs to restore all the stale nodes recorded by the bitmap lines; AW-M only restores the counter blocks and AW-L never restores any security metadata nodes although their locations have been recorded in the bitmap lines. 

STAR recovers the stale SIT nodes and counter blocks from top to down. When restoring one stale metadata node, since we don't know which counter is dirty, all the LSBs from the 8 child nodes and the corresponding counter in the father node need to be read to restore the 8 counters and MAC in this node. Then STAR caches all the MACs of metadata nodes recorded by bitmap lines, orders them by the descending addresses in a set, generates the set-MACs and reconstructs the cache-tree root. A replace attack during recovery process makes the recalculated MAC wrong. The wrong MAC is read into cache and reconstructs a wrong cache-tree root. Finally, the system verifies the recovery process by matching the recalculated cache-tree root and the cache-tree root stored on chip. If the two roots are not matched, the system recovery fails.

\iffalse
When an attacker replaces a bitmap line or a child node of a dirty node, our STAR can detect that the attack occurs and the recovery fails. But STAR doesn't know in which data node the attack occurs. The location where the single point attack occurs can't be known by Anubis.
\fi

\section{Performance Evaluation}
\label{section 6}
In this section, we first show the configurations of our simulation. We then present the experimental results and analysis.

\vspace{-0.2cm}
\subsection{Evaluation Methodology}
\label{section 5}
\vspace{-0.1cm}

\begin{table}[!htbp]
	%\vspace{5px}
	\footnotesize
	%\scriptsize
	%\tiny
	\vspace{-0.2cm}
	\caption{\label{table:configure} The configurations of the NVM system.}
	\vspace{-0.3cm}
	%\vspace{-5px}
	\begin{center}
		\begin{tabular}{|c|l|}
			\hline
			\multicolumn{2}{|c|}{\textbf{Processor}} \\
			\hline
			CPU & 4 cores, X86-64 processor, 2 GHz        \\
			\hline
			Private L1  cache & 64KB, 2-way, LRU, 64B Block              \\
			\hline
			Private L2 cache & 512KB, 8-way, LRU, 64B Block        \\
			\hline
			Shared L3 cache & 4MB, 8-way, LRU, 64B Block            \\
			\hline
			
			\multicolumn{2}{|c|}{\textbf{DDR-based PCM Main Memory}} \\
			\hline
			Capacity &  16GB     \\
			\hline
			PCM latency model &   \tabincell{c}{Read 60ns, Write 150ns}     \\
			\hline
			\multicolumn{2}{|c|}{\textbf{Secure Parameters}} \\
			\hline
			Counter Cache & 256KB, 8-Way, 64B Block, in MC           \\
			\hline
			SIT Cache & 256KB, 8-Way, 64B Block, in MC             \\
			\hline
			SIT & \tabincell{c}{10 levels, including the counter node and root, \\
				8-ary, 64B Block}             \\
			\hline
		\end{tabular}
	\end{center}
	\vspace{-8px}
	
\end{table}

To evaluate the performance of STAR, we use Gem5~\cite{BinkertBBRSBHHKSSSSVHW11} with NVMain~\cite{PorembaZ015} to model the system. NVMain is a cycle-accurate main memory simulator for emerging NVM technologies. As illustrated in Table~\ref{table:configure}, we simulate a 4 core X86 processor. The NVM system contains 32KB L1 data and instruction caches, 512KB L2 and 4MB shared L3 caches. Both the counter cache and SIT cache are 256KB, which serve as metadata cache managed by memory controller~\cite{zuo2019supermem}. We use 16GB PCM-based main memory with the read/write latency of 60ns/150ns like existing schemes~\cite{ZubairA19,AwadYSNZ19,YangLCMS19}. We use 8 representative applications from SPEC 2006 benchmark suite~\cite{Henning06} to simulate 5 billion instructions for each application, and five persistent workloads which are widely used in existing schemes in persistent memory~\cite{CoburnCAGGJS11,RenZKCWM15,KolliRDSPLCW16,KolliGSDCNW17,LiuKRK18,Zuo018}, to evaluate the performance of STAR. The five persistent workloads are as followings.

%i.e., array, btree, queue, hash, rbtree,
\begin{enumerate}
	\item{\emph{\textbf{array.}}} Initializing a 1GB array and then randomly swapping entries.
	\item{\emph{\textbf{btree.}}} Inserting random key-value items into a 1GB B-tree based key-vaule store.
	\item{\emph{\textbf{hash.}}} Inserting random key-value items into a 1GB hash table based key-value store.
	\item{\emph{\textbf{queue.}}} Randomly enqueueing and dequeueing entries in a 1GB queue.
	\item{\emph{\textbf{rbtree.}}} Inserting random key-value items into a 1GB red-black tree.
\end{enumerate}

To comprehensively examine the performance of our proposed STAR, we evaluate the following schemes for comparisons.

\begin{itemize}
	\item A write back metadata cache (WB). It uses an ideal write-back metadata cache in which only the evicted data from metadata cache is flushed into NVM. Since not all changed metadata are persistent, the WB scheme doesn't support recovery after system crashes.
	\item A strict persistence scheme (Strict Persistence): Strict Persistence scheme persists all the changed nodes from the modified counter block up to the root of SIT.
	\item Anubis for SGX Integrity Tree (ASIT) scheme (Anubis): ASIT scheme is designed for SIT in Anubis. An ST block, in which ASIT records the address of dirty metadata, correct counters and MAC, is written into NVM with each memory write.
	\item AW-L: The AW start-level of AW-L is the user data level. AW-L incurs 2 writes when writing user data and only needs to verify the correctness of recovery process without restoring stale metadata.
	\item AW-M: The AW start-level of AW-M is the counter block level. AW-M incurs 2 writes when writing counter blocks and needs to restore the stale counter blocks.
	\item AW-H: The AW start-level of AW-H is the (N-1)th level. AW-H needs to restore all the stale metadata, including the counter blocks and SIT nodes.
\end{itemize}

Our STAR doesn't switch the AW start-level during runtime. The different configurations of the AW approach meet different requirements of write and recovery overheads.

Note that since Osiris and Triad-NVM can't be used to recover the counter blocks and integrity tree nodes in SIT persistent memory, we don't compare our STAR with them.

\begin{figure}[t]
	%\vspace{-0.3cm}
	\centering
	\includegraphics[width=0.45\textwidth]{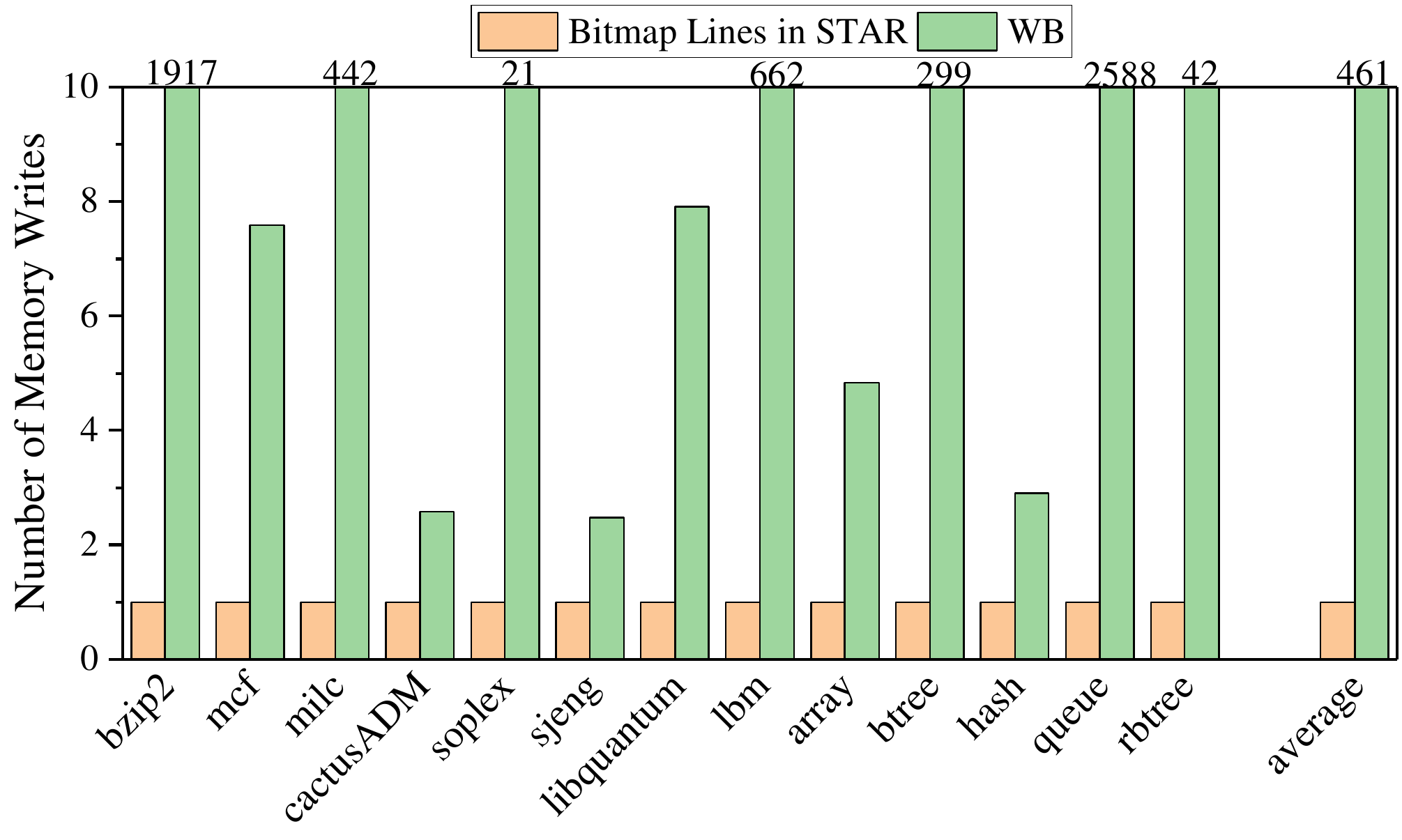}
	\caption{The number of writes of Bitmap Lines in STAR compared with that of WB (normalized to Bitmap Lines).}
	\label{bitmapline}
	\vspace{-0.45cm}
\end{figure}

\begin{figure}[h]
	%\vspace{-0.3cm}
	\centering
	\includegraphics[width=0.45\textwidth]{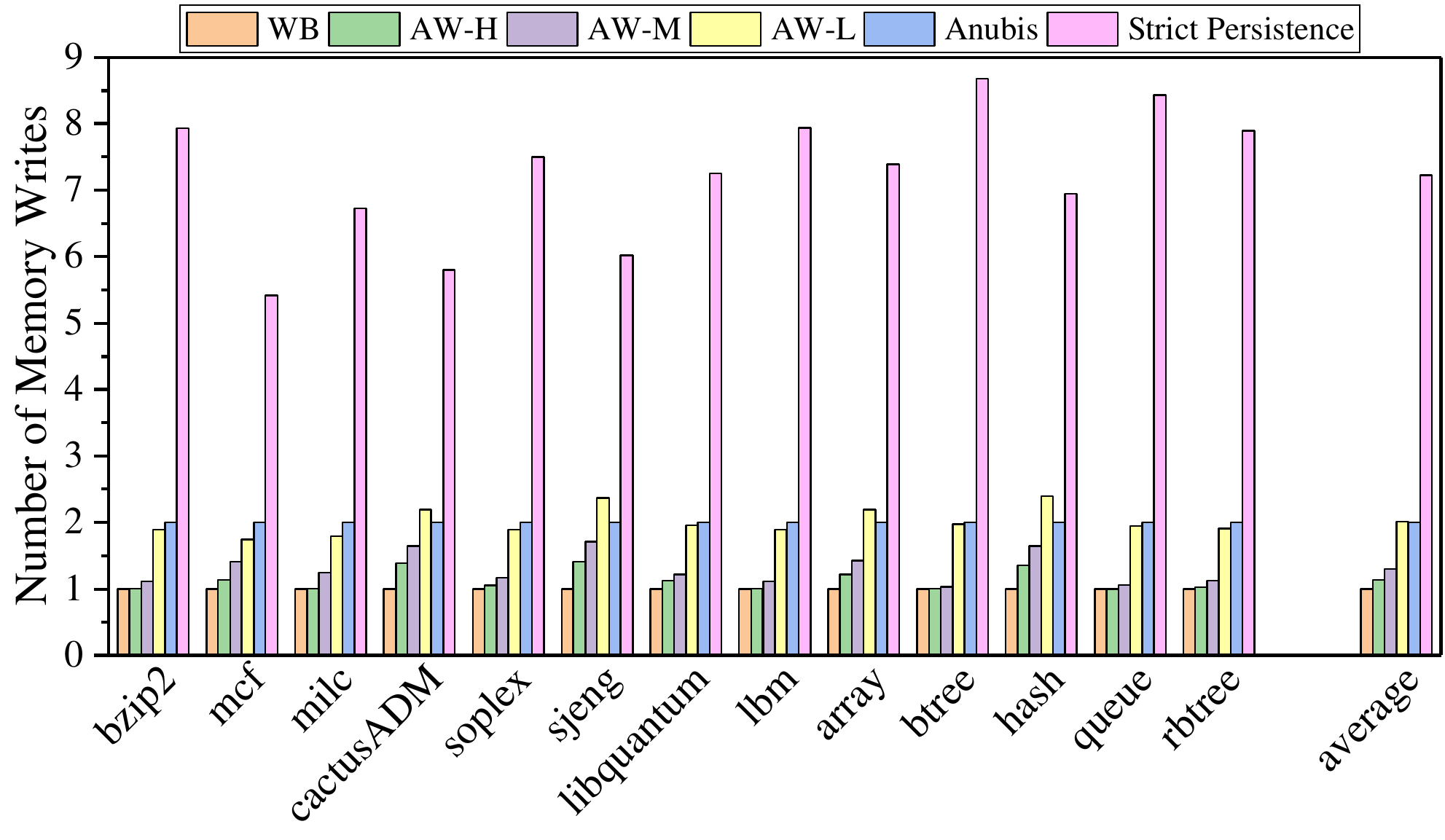}
	\caption{The number of all memory writes in different schemes (normalized to WB).}
	\label{overall_writes}
	\vspace{-0.3cm}
\end{figure}

\vspace{-0.1cm}
\subsection {Performance Results and Analysis}
\vspace{-0.1cm}
The overall write overhead of STAR consists of bitmap lines writes, ahead writes and normal memory line writes. We present the overhead of the bitmap lines writes and all write overheads of STAR compared with other schemes.

Fig.~\ref{bitmapline} shows the number of bitmap lines writes of STAR scheme compared with the WB scheme. The number of bitmap lines in ADR is 16, consisting of 2 L2 bitmap lines and 14 L1 bitmap lines. STAR flushes the bitmap lines to NVM since there is a bitmap line miss and one line should be evicted into RA by LRU. Fig.~\ref{bitmapline} shows that the overhead of the bitmap lines writes in STAR is negligible compared with WB. On average, the number of WB writes is 461x more than that of Bitmap lines writes. This is because one bitmap line covers 8-page metadata address (512 * 64B = 32KB). When the memory writes of applications have a high spatial locality, STAR rarely evicts a bitmap line. The most bitmap lines writes are caused by security metadata lines evicted from metadata cache since they have lower spatial locality than user data. Fig.~\ref{bitmapline} shows that in different workloads, the numbers of bitmap lines writes become different, which depend on two factors, i.e., the number and the spatial locality of this workload writing. The more user data is written and the higher the spatial locality of memory writes is, the bitmap lines consumes less write overhead compared with WB.

Fig~\ref{overall_writes} shows the overall number of memory writes for different schemes. The numbers of memory writes in Anubis/strict persistence scheme are the sum of traditional memory writes(i.e., the writes in WB scheme) and ST block writes/nodes in a branch of tree writes in SIT (the depth of SIT in 16GB persistent memory is 10, including counter block and root). Note that the number of strict persistence scheme writes is less than 9 times the WB, since the root is not flushed, and the tree nodes also need to be evicted in the WB scheme according to cache replacement policy. Compared with the baseline WB scheme, the number of memory writes of AW-H/AW-M/AW-L is 1.13x/1.29x/2.01x, while Anubis writes 2x memory lines than WB. The extra memory writes in AW-H are introduced by writing bitmap lines. Unlike AW-H, AW-M and AW-L have more memory writes caused by memory line ahead writes when writing nodes in the start-level. For AW-L, flushing user data will incur its counter block ahead writing, and for AW-M, only every counter block writing incurs its father node ahead writing. Since the frequency of flushing user data is higher than that of flushing counter blocks, AW-L writes more extra memory lines than AW-M and AW-H. Moreover, AW-L writes a similar number of memory lines with Anubis, but consumes less recovery time as shown in Section~\ref{recoverytime}.

%\begin{figure}[t]
%\vspace{-0.3cm}
%\centering
%\includegraphics[width=0.45\textwidth]{experiment/bitmap_hitratio.pdf}
%\caption{The hit ratio of different number of bitmap lines placed in ADR (2, 4, 8, 16 and 32). }
%\label{bitmap_hitratio}
%\vspace{-0.45cm}
%\end{figure}

\vspace{-0.2cm}
\subsection {Sensitivity to the Number of Bitmap Lines in ADR}
\vspace{-0.1cm}
A bitmap line in ADR can cover 8 memory pages since one bitmap line contains 512 bits and one bit represents one memory line. More bitmap lines in ADR cover more metadata space, thus increasing the hit ratio of the bitmap lines. Table~\ref{table:bitmap_hitratio} shows the average hit ratio of over 13 workloads among 2, 4, 8, 16 and 32 bitmap lines in ADR. The hit ratio is not too high because the system tries to access the bitmap lines only when the state of one metadata line is changed. For example, a user data is written into NVM making its corresponding counter block dirty from the clean state. The location of this counter block needs to be recorded in the bitmap lines. Next, the neighbor user data with the same counter block is written. Since the counter block has been already dirty and the state of the counter block is untouched, the system doesn't need to access the bitmap lines and record this location again. When the dirty metadata is evicted from cache, the system records its location in a bitmap line due to its state changing from dirty to the clean state. But the system will not access a bitmap line when a clean metadata is evicted. If the bitmap lines record all locations of the evicted nodes and their father nodes, no matter their states are changed or not, the hit ratio will be high, with some useless bitmap lines access. Table~\ref{table:bitmap_hitratio} shows that, the more the number of bitmap lines exists in ADR, the higher the hit ratio becomes. A higher hit ratio causes less number of bitmap lines written. Considering that the ADR region on chip is expensive and the improvement of hit ratio decreases with more bitmap lines, we choose to place 16 bitmap lines in ADR.

\begin{table}[t]
	%\vspace{5px}
	\footnotesize
	%\scriptsize
	%\tiny
	\vspace{-0.2cm}
	\caption{\label{table:bitmap_hitratio} The hit ratios of different numbers of bitmap lines placed in ADR (2, 4, 8, 16 and 32).}
	\vspace{-0.3cm}
	%	\vspace{-5px}
	\begin{center}
		\begin{tabular}{|c|c|c|c|c|c|}
			\hline
			Bitmap Lines & 2 & 4 & 8 & 16 & 32\\
			\hline
			Hit Ratio & 32.85\% & 47.44\% & 64.37\% & 74.75\% & 82.19\% \\
			\hline
		\end{tabular}
	\end{center}
	%	\vspace{-7px}
	\vspace{-0.3cm}
\end{table}

\begin{figure}[t]
	%\vspace{-0.3cm}
	\centering
	\includegraphics[width=0.45\textwidth]{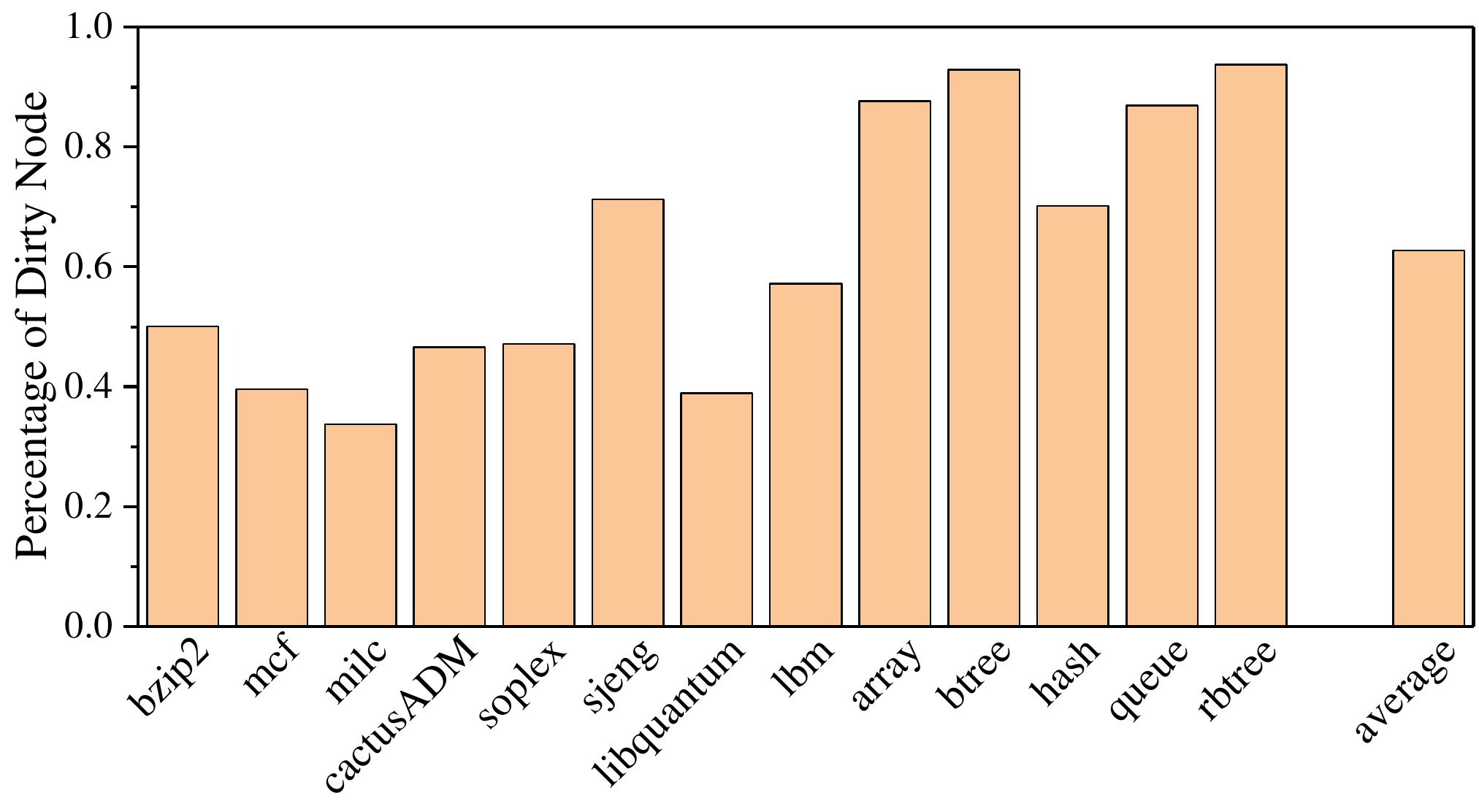}
	\caption{The percentage of dirty nodes in metadata cache on different workloads.}
	\label{dirty_ratio}
	\vspace{-0.45cm}
\end{figure}

\begin{figure}[t]
	%\vspace{-0.3cm}
	\centering
	\includegraphics[width=0.45\textwidth]{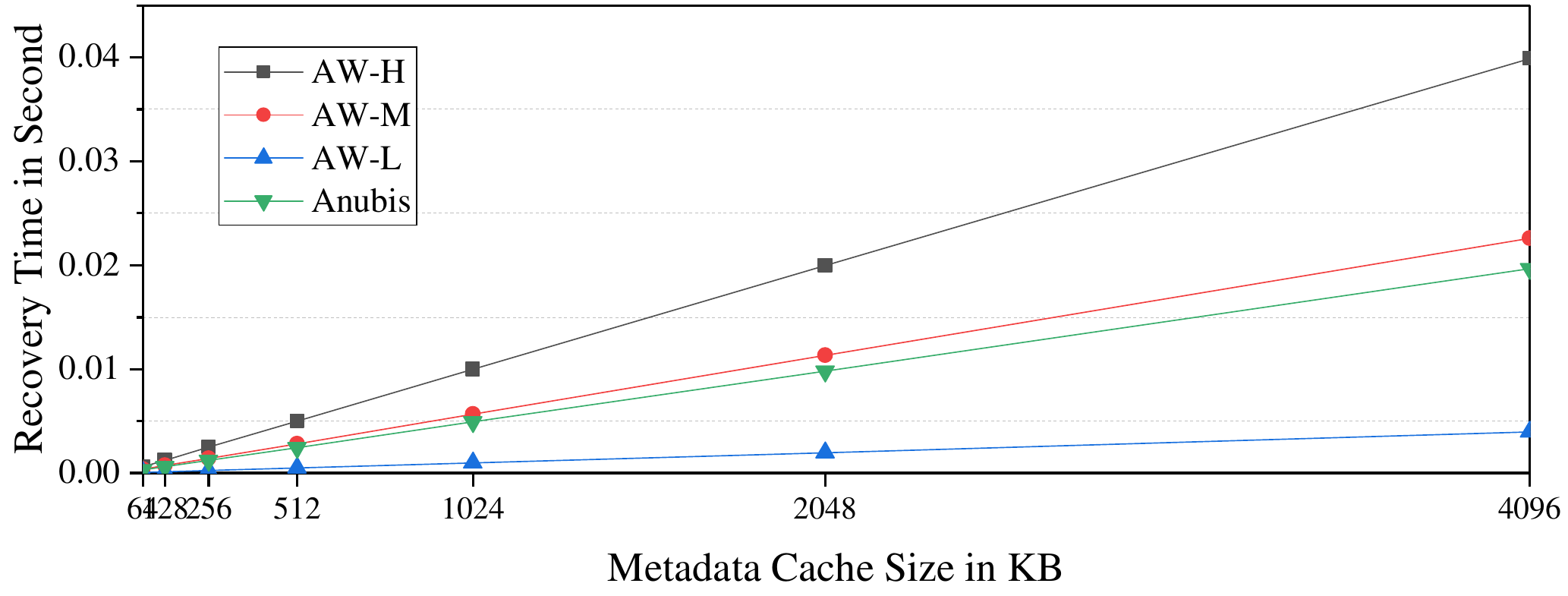}
	\caption{The recovery time in different sizes of metadata cache for three configurations of STAR and Anubis.}
	\label{recovery_time}
	\vspace{-0.45cm}
\end{figure}

\vspace{-0.3cm}
\subsection {Recovery time}
\label{recoverytime}
\vspace{-0.1cm}
To recover the stale node, STAR first reads the non-zero L1 bitmap lines according to the multi-layer index to obtain the locations of stale metadata node. STAR further recovers the stale nodes by reading its father node and 8 child nodes to restore the counters and MAC in the stale node. Finally, the cache-tree is reconstructed to verify the correctness of the recovery process. Like Anubis and Osiris~\cite{ZubairA19,YeHA18}, we assume fetching one node from NVM to cache and updating it would consume 100ns. The recovery time consists of fetching metadata nodes from memory and reconstructing the cache-tree. The latency of reading father node and child nodes dominates the recovery time.

Fig.~\ref{dirty_ratio} shows the percentage of dirty nodes in metadata cache. When system crashes occur, these dirty nodes need to be restored. STAR needs to read 62\% nodes of cache instead of 100\% read in Anubis. STAR further provides three different configurations for different write overheads and recovery time, i.e., AW-L, AW-M and AW-H. AW-L doesn't need to restore metadata node, and only needs to verify the correctness of recovery process, thus introducing short recovery time. For verifying the correctness of recovery process, AW-L needs to read the MAC of dirty node and reconstruct the cache-tree. In addition to the verification of recovery process, AW-M needs to restore the stale counter blocks and AW-H has to restore all the stale metadata, including counter blocks and SIT nodes.

Fig.~\ref{recovery_time} shows the recovery time after system crashes of different schemes. For a 4MB metadata cache, AW-H/AW-M/AW-L needs 0.039s/0.023s/0.004s to recover the stale security metadata while Anubis needs 0.020s. Compared with Anubis, AW-H needs 1.99x recovery time, AW-M needs 1.15x and AW-L needs 0.20x. During the recovery process, Anubis reads the ST blocks and their number is equal to that of metadata cache lines. According to the ST blocks, all the metadata in metadata cache and the father nodes of these metadata are read to be restored by Anubis, no matter the metadata is stale or not in NVM. Therefore, for a 4MB metadata cache, Anubis reads 196,608 lines from NVM (3x than the number of lines in 4MB metadata cache) during recovery process. Unlike Anubis, AW-L reads the negligible bitmap lines (about 150 lines in our experiments) and dirty metadata in metadata cache (about 62\% number of metadata cache lines) to verify the recovery process. AW-M and AW-H read more memory lines for recovery due to restoring stale nodes. AW-M restores the stale counter blocks while AW-H restores all the stale metadata nodes. Restoring each dirty metadata node needs to read 10 related nodes, including 1 dirty node to be restored, 1 father node and 8 child nodes.

\section{Related Work}
\label{section 7}
In this section, we discuss the prior works related with our STAR.

\textbf{Recovery in NVM.} To use the data persisted in NVM after system crashes, a secure persistent memory system needs to recover the security metadata. Osiris~\cite{YeHA18} recovers the counter blocks by retrying counter and leverages error-correction codes to verify the correctness of recovered counter blocks. Since Osiris writes a counter block into NVM when one counter in this block has been increased by N times, Osiris has fewer writes than conventional write back schemes. cc-NVM~\cite{YangLCMS19} caches the flushed counter blocks in write pending queue with a battery in an epoch and only flushes these blocks at the end of one epoch. cc-NVM also retries the counter to obtain the correct one. Unlike Osiris, cc-NVM uses MAC to verify which counter is correct. To recover a merkle tree, Triad-NVM~\cite{AwadYSNZ19} flushes the n lowest level tree nodes and counter blocks with the user data writes. On recovery, Triad-NVM reconstructs the whole tree from the flushed tree nodes instead of user data. Anubis~\cite{ZubairA19} provides a fast recovery scheme for merkle tree and SGX integrity tree by recording the addresses of the changed metadata in a shadow table block and flushing the shadow table block with user data. Anubis only recovers the cached nodes before crashes according to shadow table blocks and reduces the recovery time.
\iffalse
\textbf{Security metadata organization.} To reduce the access latency and space overhead of integrity tree, existing schemes propose many variants of SIT. VAULT~\cite{TaassoriSB18} reduces the depth and space overhead of SIT by storing more than 8 counters in one node, e.g., 64 counters in the leaf node, 32 counters in the level-2 node and 16 counters in level-3 and upper-level nodes. VAULT reduces the paging overhead by increasing the size of Enclave Page Cache to match that of the physical memory. Based on VAULT, Morphable Counters~\cite{SaileshwarNREJQ18} observes that when one counter overflows, either less than a quarter of counters or all the counters are used. Morphable Counters provides a scalable solution to store 128 counters in one node and further reduce the height of the tree. Synergy~\cite{SaileshwarNREQ18} places the MAC inside the ECC chip in a 9-chip ECC-DIMMs and demonstrates that MAC can be used to detect not only data tempering but also memory errors. With the aid of Synergy, a system can avoid a separate memory access for MAC and reduces the overall memory traffic.
\fi

\textbf{Secure NVM.} Unlike DRAM, NVM suffers from the data remanence vulnerability and limited lifetime. Ensuring data confidentiality with low write overhead in NVM is necessary. DEUCE~\cite{YoungNQ15} proposes a dual-counter scheme. In one epoch, DEUCE uses the old counter to encrypt the untouched words and new counter to encrypt the changed words in a cache line, which reduces the write traffic since the untouched words needn't to be written by executing DCW~\cite{YangLKCLY07} and FNW~\cite{ChoL09} techniques. Based on DEUCE, SECRET~\cite{SwamiRM16} further reduces the zero-content words writes in NVM by flushing a zero-flag. Silent Shredder~\cite{AwadMHSH16} observes that zeroing out physical pages before mapping to processor consumes a large percentage of memory writes. They repurpose initialization vectors of counter mode encryption to eliminate the data shredding writes. SuperMem~\cite{zuo2019supermem} uses a write-through counter cache to ensure the counter crash consistency and proposes counter write coalescing (CWC) scheme to reduce the number of counter writes.

Unlike existing schemes, STAR reduces the number of the SIT nodes and counter blocks that need to be recovered to achieve low recovery and write overheads. Moreover, Our STAR is orthogonal with SuperMem, and the CWC scheme can be used in the AW-L to further reduce the number of memory writes. 

\section{CONCLUSION}
\label{section 8}
This paper proposes STAR to reduce the high recovery time and write overhead of recovering the SGX integrity tree nodes and counter blocks in the secure non-volatile memories. STAR leverages Ahead Write (AW) scheme to provide the consistency between the nodes in metadata cache and their copies in NVM. Furthermore, to efficiently restore the dirty metadata and verify the recovery process, STAR judiciously exploits the unused space in MAC in one node to store the LSBs of the corresponding counter in the father node, and leverages bitmap lines in ADR to show the locations of stale metadata nodes. Moreover, a cache-tree is constructed to ensure the correctness of the recovery process and a multi-layer index is introduced to speed up recovering stale nodes. Experimental results show that compared with state-of-the-art work, STAR reduces the number of memory writes by up to 87\% with a short recovery time. 

%-------------------------------------------------------------------------------
%\bibliographystyle{plain}
\bibliographystyle{IEEEtranS}
\bibliography{references}
	
	%%%%%%%%%%%%%%%%%%%%%%%%%%%%%%%%%%%%%%%%%%%%%%%%%%%%%%%%%%%%%%%%%%%%%%%%%%%%%%%%
\end{document}